# Studies on thermal degradation behavior of nano silica loaded cotton and polyester fabrics

S. B. Chaudhari[2], P. N. Patel[1], A. A. Mandot[2] and B. H. Patel[1]

*1 Department of Textile Chemistry, The Maharaja Sayajirao University of Baroda, Vadodara, Gujarat, India.*
*2 Department of Textile Engineering, The Maharaja Sayajirao University of Baroda, Vadodara, Gujarat, India.*

**Abstract** In this work cotton and polyester fabric were coated with silica nano particles using pad-dry-cure method. The prepared composite fabrics were analyzed in terms of change in their thermal degradation properties. The oxidative degradation characteristic of control and treated samples were studied using Thermo Gravimetric technique. The mass loss profile of untreated fabric compared with treated fabric. The results indicate that the thermal degradation rate was lower in treated sample as compare to control fabrics.

## 1. Introduction

Nanotechnology seeks to provide and apply knowledge of the behavior of objects in the nanometer (nm) size range to the assembly of complex structures for use in a variety of practical applications. The tiniest substances promise to transform industry and create a huge market. In chemicals, cosmetics, pharmaceuticals, technology and textiles, businesses are doing research and manufacturing products based on nanotechnology, which uses bits of matter measured in billionths of a meter [1-3].

Thermal properties are the properties of materials that changes with temperature. They are studied by thermal analysis techniques, which include DSC, TGA, DTA, dielectric thermal analysis, etc. Generally, the incorporation of nanometer-sized inorganic particles into the polymer matrix can enhance thermal stability by acting as a superior insulator and mass transport barrier to the volatile products generated during decomposition [3-7].

---

1)  Email: s.b.chaudhari-ted@msubaroda.ac.in

Nanoparticles can consist of various elements and compounds. The size of the molecules is the sole criterion for inclusion in the category of nanoparticles. Some of the inorganic nanosize particles currently available for use are $TiO_2$, $ZnO$, $Fe_2O_3$, $Ag$, $Pt$, $SiO_2$ etc. Among these nano particles $SiO_2$ nanoparticle are of most important [8-10]. As silica is most commonly found in nature as sand or quartz as well as in the cell walls of diatoms. Silica is manufactured in several forms including fused quartz, crystal, fumed silica, colloidal silica, silica gel, and aerogel. At present, there are several techniques reported in literatures for synthesizing nanoparticles [11-14], however, wet-chemical technique is widely used in the field of material science and ceramic engineering [15-18]. The coating of nano-particles on fabrics will not affect their breathability or feel. Therefore, the use of nanotechnology in textile is increasing day by day.

In this work attempt has been made to apply $SiO_2$ nano particles to cotton and polyester fabrics using pad-dry-cure technique. The in house synthesized $SiO_2$ nano particles have been elementally characterized EDX. The surface morphology of treated fabrics has been observed by SEM technique. Finally, the thermal degradation behavior of $SiO_2$ nano treated cotton and polyester fabric have been studied by Thermo Gravimetric technique.

## 2. Materials and Methods

### 2.1 Materials

*2.1.1 Fabric*

Mill scoured and bleached cotton and fabrics with specification as given in Table 1 were procured from local market and used for the study. The procured fabric was further thoroughly washed, neutralized and air dried.

**Table 1** Specification of cotton and polyester fabric

| Sample | Material Specification | | | | | | |
|---|---|---|---|---|---|---|---|
| | Count/Denier | | Ends /inch | Pick/inch | Weave | Wt.gm/sq.m. | Thickness (mm) |
| | Warp | Weft | | | | | |
| Cotton | 35s | 31s | 116 | 88 | Plain | 118.1 | 0.23 |
| Polyester | 128d | 146d | 90 | 72 | Plain | 109.7 | 0.21 |

*2.1.2 Chemicals*

In-house synthesized nano $SiO_2$ nanoparticles[12] and poly acryl amide ($CH_2CHCONH_2$, MW 71.08) of analytical grade purity was procured from SuLab reagents.

**2.2 Experimental methods**

*2.2.1 Treatment of cotton and polyester with silica nano-particles*

The coating solutions containing nano silica particle were prepared using 1gpl, 2.5 gpl, and 5 gpl concentration, i.e. for 1 gpl solution, 0.1 gm nano particle was added with 5 gm Lissapol L surfactant and 10gm polyacrylamide binder. The mixture was then stirred using magnetic stirrer at 250 rpm for 30 minutes at 60°C temperature. Likewise all concentration solution was prepared.

The padding liquor was applied to the cotton and polyester fabric samples (size : 40 cm X 30 cm) by dipping them in the dispersion for 10 min and then padded on an automatic padding mangle machine, which was running at a speed of 15 rpm with a pressure of 1.75 Kg/cm$^2$ using 2-dip-2-nip padding sequence at 70% expression. The padded substrates were air dried and finally cured for 3 min at 120 ˚C and 130 ˚C respectively for cotton and polyester fabric.

**2.3 Testing and Analysis**

*2.3.1 Fabric characterization Techniques*

The surface morphology of the nano silica / cotton composite fabric was observed on scanning electron microscope (SEM) instrument (Model JSM5610LV, version 1.0, Jeol, Japan) and the presence of silica in composite fabric was confirmed on scanning electron microscope using Oxford-Inca software.

*2.3.2 Determination of thermal properties of fabrics*

Thermogravimetry (TGA) was carried out for control and treated cotton and polyester samples by using a Shimadzu TGA-50 thermal analyzer. The samples were heated from ambient temperature to 500$^o$C with 10$^o$C/min rate in normal atmosphere. The thermograms associated with TG for control and treated samples were obtained from the instrument output. The mass loss degradation at onset temperature were obtained from there thermograms for both the cotton and polyester treated as well as for untreated sample.

## 3. Results and Discussion

Fig. 1 shows the SEM micrographs of polyester and cotton fabrics loaded with silica nanoparticles distributed uniformly on fabric surface. The average diameter of in-house prepared nano particle as seen from the Fig. 1 was ~180 nm. The elemental analysis of silica nano particle was performed on scanning electron microscope using Oxford-Inca software. The presence of silica was confirmed by the elemental analysis (Table 2); also the presence of oxygen indicates the silica is in the form of oxide or dioxide.

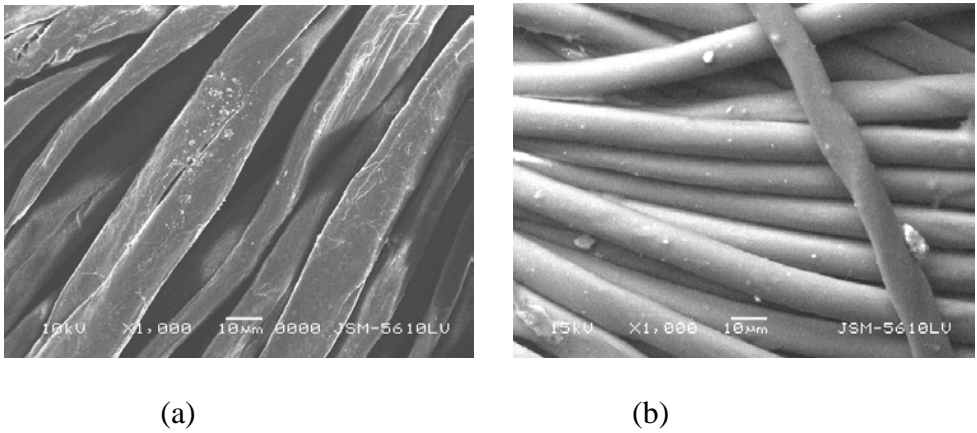

(a)                              (b)

Fig. 1. Nano silica loaded SEM images of (a) Cotton and (b) Polyester

**Table 2** Elemental analysis of silica nano coated cotton and polyester fabric

| Element | Weight% | Atomic% |
|---------|---------|---------|
| O K     | 51.03   | 61.90   |
| F K     | 12.91   | 13.18   |
| Si K    | 36.06   | 24.92   |
| Total   | 100     |         |

### 3.1 Effect of nano silica on thermal behavior of fabric

Cotton and polyester fabric were treated with silica nano particle using pad-dry-cure technique. The oxidative degradation characteristic of control and treated samples were studied by using Thermo Gravimetric technique. The mass loss profile of control fabric compared with treated fabric. The cotton cellulose normally decomposes below $300^{o}C$ and under dehydration, depolymerization and oxidation, release of $CO$, $CO_2$ and carbonaceous residual char result. Tar and leaveglucosan are formed at around $300^{o}C$. Polyester normally melts and flows under the influence of the temperature at above $260^{o}C$. In case of polyester, the thermal decomposition is initiated by scission of an alkyl-oxygen bond and the material decomposes via the formation of cyclic or open chain oligomers with olefinic or carboxylic end group at above $510^{o}C$. The thermal degradation is a combination of both physical and chemical process that involves decomposition and oxidation and depends upon the activation energy (E).

The plots of thermal degradation of control and treated cotton and polyester fabrics are given in Fig. 2. The % reduction in weight is observed during the thermal decomposition and the measured values from TGA curve are given in Table 3. The TGA curve in Fig. 2 shows thermal degradation of control and treated cotton fabric. The TGA curve is analyzed in three steps to study the mass loss. In initial step, the mass loss of control sample is of 6% and that of treated sample is 5.6% observed at the temperature range of $25-150^{o}C$.

**Table 3** Thermal degradation behaviors of cotton and polyester fabric

| Temp [°C] | % Reduction in wt. | | | |
|---|---|---|---|---|
| | Cotton | | Polyester | |
| | Untreated | Treated | Untreated | Treated |
| 25 | 0 | 0 | 0 | 0 |
| 50 | 1.75115 | 1.43095 | 0.226 | 0 |
| 75 | 2.30415 | 2.03095 | 0.225 | 0 |
| 100 | 1.19816 | 0.8704 | 0 | 0.112 |
| 125 | 0.46083 | 0.38685 | 0.225 | 0.111 |
| 150 | 0.27649 | 0.19342 | 0 | 0.112 |
| 175 | 0.09217 | 0.19343 | 0.233 | 0.111 |
| 200 | 0.09217 | 0.19342 | 0.008 | 0.111 |
| 225 | 0.18433 | 0.38685 | 0 | 0.112 |
| 250 | 0.36866 | 0.96711 | 0.226 | 0.111 |
| 275 | 1.19816 | 2.70794 | 0.225 | 0.111 |
| 300 | 3.5023 | 7.25338 | 0 | 0.112 |
| 325 | 56.84996 | 6.93277 | 0.45 | 0.111 |
| 350 | 6.19152 | 45.48503 | 0.676 | 0.446 |
| 375 | 2.39631 | 2.80464 | 2.027 | 1.893 |
| 400 | 2.48848 | 3.96518 | 5.856 | 9.02 |
| 425 | 9.67742 | 5.60929 | 20.495 | 23.162 |
| 450 | 0.46083 | 5.8027 | 40.09 | 38.196 |
| 475 | 2.85714 | 4.448746 | 16.667 | 9.02 |
| 500 | 4.239632 | 2.901354 | 1.126 | 1.114 |

In second step, the mass loss of control and treated samples are 6% and 11.7% respectively, which is observed at the temperature range of 150-300°C, and in final step, it was 84.6% and 77.9% of control and treated fabrics respectively observed at the temperature range of 300-500°C. From the Table 3 the control sample looses 57% of mass at the temperature range between 300-325°C, which may be due to the de-hydration of cotton and oxidation thermal decomposition to CO, $CO_2$ and formation of carbonaceous char. The treated fabric gives maximum thermal decomposition at the temperature range of 325-350°C. The results indicate that the thermal decomposition rate is higher in control sample as compare to treated fabric.

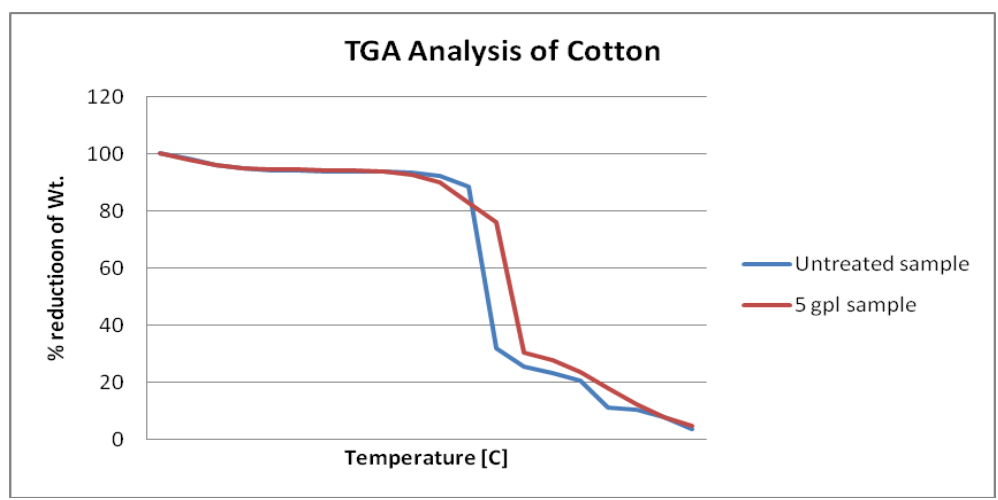

Fig. 2. Effect of temperature on weight reduction of cotton fabric

In case of polyester the TGA curve in Fig. 3 shows the thermal degradation of control and treated fabrics. In initial step the mass loss of control and treated fabrics are 0.6% and

0.34% respectively. In second step it is 0.76% and 0.67% of control and treated fabrics respectively. In third step, the mass loss is 87.38% and 82.95% of control and treated fabrics respectively. The maximum thermal degradation of polyester treated and untreated fabrics observed at 425-450°C temperature range. The results indicate that the thermal degradation rate is lower in treated sample as compare to control fabric.

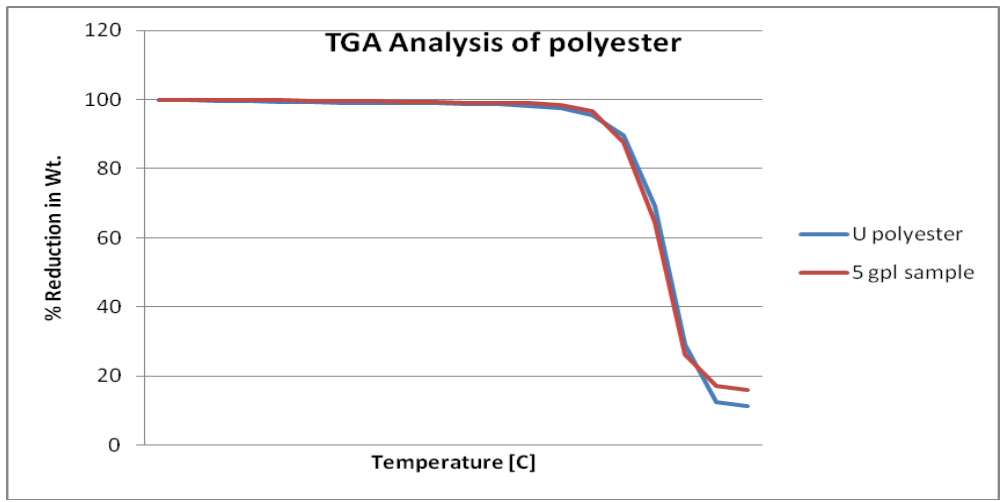

Fig. 3. Effect of temperature on weight reduction of polyester fabric

## 4. Conclusions

The in house synthesized nanosilica powder can be applied on cotton and polyester fabrics by pad-dry-cure technique. EDS results further confirm the existence of silica nano particles in their oxide form. The SEM image of treated cotton and polyester with silica indicates that the silica particles are uniformly distributed on individual fibre of fabric. The mass loss profile or thermal degradation rate with respect to temperature is lower in treated fabric compared to untreated fabric.

At last Nanotechnology holds an enormous, promising future for textile. The new concept exploited for the development of nano-finishes have opened up exciting opportunities for further research and development in this area.

**Acknowledgement**

Authors are thankful to Research and Development Cell of The Maharaja Sayajirao University of Baroda for financial assistance for this study.